\documentclass[lettersize,journal]{IEEEtran}
\usepackage{algorithmic}
\usepackage{algorithm}
\usepackage{array}
\usepackage{subcaption}
\usepackage{textcomp}
\usepackage{stfloats}
\usepackage{url}
\usepackage{xurl}
\usepackage{verbatim}
\usepackage{graphicx}
\usepackage{cite}
\usepackage{tikz}
\usepackage{gnuplottex}
\usepackage{epstopdf}
\usepackage{amsthm, amsmath,amssymb, wasysym}
\usepackage{amsfonts}
\usepackage{bm}

\usepackage{booktabs}
\usepackage{graphicx}

\usepackage{siunitx}

\usepackage{enumitem} 

\usepackage{multirow}

\usepackage{comment}
\usepackage{color}

\usepackage{xspace}

\usepackage{tcolorbox}

\usepackage[colorlinks, linkcolor=blue, urlcolor=cyan]{hyperref}

\usepackage{pifont}

\usepackage{pict2e}

\newsavebox{\ORCIDlogo}
\savebox{\ORCIDlogo}{%
\setlength{\unitlength}{\dimexpr 1em/256\relax}%
\begin{picture}(256,256)%
  \color[HTML]{A6CE39}\put(128,128){\circle*{256}}%
  \color{white}%
  \put(78.6,199.2){\circle*{20}}%
  \moveto(70.9,176,9)\lineto(86.3,176,9)\lineto(86.3,69.8)\lineto(70.9,69.8)%
  \closepath\fillpath%
  \moveto(108.9,176.9)\lineto(150.5,176.9)%
  \curveto(190.1,176.9)(207.5,148.6)(207.5 ,123.3)%
  \curveto(207.5,95,8)(186,69.7)(150.7,69.7)%
  \lineto(108.9,69.7)%
  \closepath\fillpath%
  \color[HTML]{A6CE39}%
  \moveto(124.3,83.6)\lineto(148.8,83.6)%
  \curveto(183.7,83.6)(191.7,110.1)(191.7,123.3)%
  \curveto(191.7,144.8)(178,163)(148,163)%
  \lineto(124.3,163)%
  \closepath\fillpath%
\end{picture}%
}

\newcommand\orcidicon[1]{\href{https://orcid.org/#1}{\usebox{\ORCIDlogo}}}

\newcommand{\attack}{{WFCAT}\xspace}

\usepackage{amsmath} 
\newcommand{\uptrend}[1]{%
  $_{\textcolor{red}{\blacktriangle#1}}$%
}

\newcommand{\downtrend}[1]{%
  $_{\textcolor{blue}{\blacktriangledown#1}}$%
}

\begin{document}

\title{\attack: Augmenting Website Fingerprinting with Channel-wise Attention on Timing Features}

\author{
Jiajun GONG~\orcidicon{0000-0001-9906-8838}, 
Wei Cai~\orcidicon{0009-0005-8025-9323},
Siyuan Liang~\orcidicon{0000-0002-6154-0233},
Zhong Guan~\orcidicon{0009-0002-2072-894X},
Tao Wang~\orcidicon{0000-0003-3886-0420},
and Ee-Chien Chang~\orcidicon{0000-0003-4613-0866}
       
\thanks{
Jiajun Gong, Siyuan Liang, and Ee-Chien Chang are with the School of Computing, National University of Singapore, Singapore (email: jgongacaaron@gmail.com; pandaliang521@gmail.com; changec@comp.nus.edu.sg).
Wei Cai is with the Network Connection Security Department, Zhongguancun Lab, China (email: caiwei@zgclab.edu.cn).
Zhong Guan is with the Institute of Information Engineering, Chinese Academy of Sciences (email: guanzhong@iie.ac.cn)
Tao Wang is with the School of Computing Science, Simon Fraser University, Canada (email: taowang@sfu.ca). 
Ee-Chien Chang is the corresponding author.
}

}

\markboth{IEEE TRANSACTIONS ON INFORMATION FORENSICS AND SECURITY,~Vol.~*, No.~*, December~2024}%
{Shell \MakeLowercase{\textit{et al.}}: A Sample Article Using IEEEtran.cls for IEEE Journals} 

\IEEEpubid{0000--0000/00\$00.00~\copyright~2024 IEEE}  

\maketitle

\begin{abstract}
Website Fingerprinting (WF) aims to deanonymize users on the Tor network by analyzing encrypted network traffic. 
Recent deep-learning-based attacks show high accuracy on undefended traces.
However, they struggle against modern defenses that use tactics like injecting dummy packets and delaying real packets, which significantly degrade classification performance.
Our analysis reveals that current attacks inadequately leverage the timing information inherent in traffic traces, which persists as a source of leakage even under robust defenses. 
Addressing this shortfall, we introduce a novel feature representation named the Inter-Arrival Time (IAT) histogram, which quantifies the frequencies of packet inter-arrival times across predetermined time slots. 
Complementing this feature, we propose a new CNN-based attack, \attack, enhanced with two innovative architectural blocks designed to optimally extract and utilize timing information. 
Our approach is to use kernels of varying sizes to capture multi-scale features, which are then integrated using a weighted sum across all feature channels to enhance the model's efficacy in identifying temporal patterns.
Our experiments validate that \attack substantially outperforms existing methods on defended traces in both closed- and open-world scenarios. 
Notably, \attack achieves over 59\% accuracy against Surakav, a recently developed robust defense, marking an improvement of over 28\% and 48\% against the state-of-the-art attacks RF and Tik-Tok, respectively, in the closed-world scenario.
\end{abstract}


\section{Introduction} 

With millions of daily users, Tor~\cite{DingledineMS04} stands as one of the most widely adopted technologies for safeguarding online privacy.
Tor achieves this by establishing an encrypted circuit that routes network traffic through three nodes worldwide, effectively anonymizing the user’s identity and location from the web pages they visit and any on-path eavesdroppers.
Despite its robust design, Tor remains vulnerable to a class of traffic analysis attacks known as Website Fingerprinting (WF)~\cite{SirinamIJW18, RimmerPJGJ18Automated, Hayes16kfin, Panchenko16Web, mohammadTik19, bhat2019var, shen23subverting, WangCNJG14}.
In these attacks, a local eavesdropper passively gathers side-channel information (e.g., packet sizes and inter-packet timings) from the network traffic between the victim and the entry node (i.e., the first node of a Tor circuit).
Their goal is to figure out the destination of the victim, thus breaking Tor's anonymity guarantee.

Modern WF attacks utilize deep learning models to automatically extract useful features from the raw traffic sequence, which consists of either raw packet directions~\cite{RimmerPJGJ18Automated, SirinamIJW18, JinLLS23Transformer, DengYLZLXXW23Robust} or packet timestamps~\cite{mohammadTik19, bhat2019var}.
A recent attack, RF~\cite{shen23subverting}, proposes a new 2-dimensional input feature called TAM, which aggregates the number of packets in fixed-sized time windows for incoming and outgoing packets.
TAM focuses on capturing coarse-grained packet statistics over time, making local perturbations of a defense less effective.
They show that such an input representation helps the model learn more robust features, thereby enhancing performance against a variety of existing defenses.
However, we observe that RF can still be weakened by stronger regularized defenses, which involve padding and globally delaying real packets.

We identify two design weaknesses in the existing works.
First, existing attacks do not fully exploit the use of timing information.
For example, raw timestamps used by Tik-Tok~\cite{mohammadTik19} and VarCNN~\cite{bhat2019var} can be easily perturbed under regularized defenses.
The TAM representation used by RF~\cite{DengYLZLXXW23Robust} does not use the timing information of all packets within a single time window, causing information loss.
Second, existing attacks apply conventional CNN architectures to extract features, which are not tailored for attacking defenses.
They use single small kernel sizes, which may not capture global information effectively.
Additionally, these attacks treat features from different channels as equally important, which may hinder the model’s learning process for noisy traces.

In this paper, we propose a new attack called \attack, which can significantly undermine all existing defenses.
The core idea is to enhance model learning with both better input representation and better model architecture.
We first develop an effective trace representation, involving both volume and timing information, to better capture the distinctive features of the trace.
Then, we develop a new CNN-based backbone to learn invariant features against noise to facilitate the final prediction.

\IEEEpubidadjcol 

Our contribution can be summarized as follows. 
\begin{itemize}[itemsep=5pt]
    \item We propose a multi-dimensional trace representation focusing on the timing characteristics of the packets. 
     In general, we divide the loading timeline into fixed time windows and compute a histogram for the packets within each window. 
     Specifically, we bin the packets according to their inter-arrival times and record the counts in each bin. 
     This representation preserves the timing information carried by different types of packets while being more robust against perturbations compared to raw timestamps.
    
    \item Alongside our proposed trace representation, we introduce a novel CNN block. 
    We employ multiple kernels of different sizes to capture features at various scales and then fuse all the features using different weights learned during the training process. 
    This approach ensures that highly informative features contribute more to the final prediction, making our attack robust even against defenses.
    
    \item We conduct extensive experiments on our new datasets collected on the live Tor network. 
    Experiments show that our attack achieves over 59\% accuracy against Surakav~\cite{gong2022surakav}, the state-of-the-art defense, elevating the accuracy by 28\% and 44\% compared with RF~\cite{DengYLZLXXW23Robust} and Tik-Tok~\cite{mohammadTik19}, respectively. 
\end{itemize}

\textbf{Roadmap.}
The rest of the paper is organized as follows.
In Section~\ref{sec:threat-model}, we define the threat model, outlining the capabilities and goals of both attackers and defenders in the context of website fingerprinting. 
Section~\ref{sec:related-work} reviews state-of-the-art attacks and defenses, highlighting the gaps that our work addresses. 
Section~\ref{sec:attack-design} presents the architecture and mechanisms of \attack, including its novel trace representation and feature extraction modules. 
Section~\ref{sec:attack-evaluation} provides a comprehensive evaluation of \attack's performance against multiple defenses, demonstrating its robustness and effectiveness. 
Finally, Sections~\ref{sec:discussion} and \ref{sec:conclusion} discuss relevant issues, conclude the paper, and outline future directions for advancing website fingerprinting attacks and defenses.

\section{Threat Model}
\label{sec:threat-model}

\begin{figure}
    \centering
    \includegraphics[width=\linewidth]{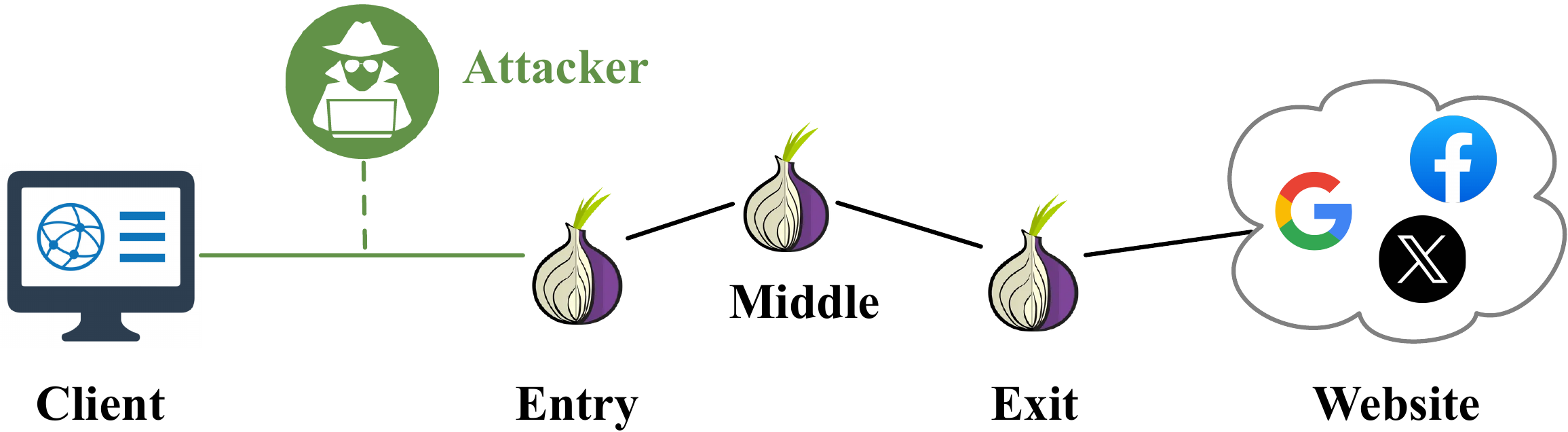}
    \caption{WF attack model.}
    \label{fig:threat-model}
\end{figure}

\textbf{Attack model.}
The Website Fingerprinting (WF) attack model is shown in Figure~\ref{fig:threat-model}. 
In this scenario, we assume the victim uses Tor to browse web pages, aiming to protect her privacy.
Within the Tor network, each packet traverses three different nodes (entry, middle, and exit) before reaching its final destination. 
The WF attacker is considered a \textit{local} eavesdropper positioned between the victim and the entry node.
The attacker knows the victim’s IP address and seeks to identify the destination page being loaded.
We assume the attacker is \textit{passive}, merely observing network patterns without altering them.
Specifically, the attacker does not drop, modify, or delay any packets, nor does he attempt to break the encryption.
Potential WF attackers include the administrator of the local network, the Internet Service Provider, or the entry node itself.

\textbf{Attack scenarios.}
Website Fingerprinting can be regarded as a classification problem where a network trace is mapped to a website label. 
To launch a WF attack, the attacker will train a classifier using pre-collected training data,
then use this trained model to predict on a given trace. 

Same as prior work~\cite{SirinamIJW18, DengYLZLXXW23Robust, bhat2019var, mohammadTik19, RimmerPJGJ18Automated}, we consider both closed- and open-world scenarios for WF attacks. 
In the closed-world scenario, the attacker \textit{monitors} a set of $N$ web pages. 
The victim is assumed to only visit these monitored sites. 
The attacker can retrieve a few network traces for each of the monitored sites (by loading these pages himself) as training samples to train the model and classify the victim's traces. 
In the open-world scenario, the victim may visit not only these $N$ monitored web pages but also other \textit{non-monitored} web pages. 
The attacker aims to predict whether a given trace belongs to a specific monitored or non-monitored website. 
In this scenario, the attacker can collect a few non-monitored training samples to include in the training source. 
Note that the non-monitored samples may only be collected from web pages that the victim never visits. 

\textbf{Defense model.}
To defend against potential WF attackers, the victim may apply an existing defense mechanism (e.g., FRONT~\cite{GongW20}, RegulaTor~\cite{HollandH22}, and Surakav~\cite{gong2022surakav}) to protect herself.
A cooperating node, optimally the middle node in the Tor network, will help obfuscate the bi-directional traffic with the client by delaying real packets or injecting dummy packets in real time. 
Consequently, the attacker can only observe the modified traffic, and due to encryption, cannot differentiate between real and dummy packets. 
We assume that the attacker is aware of the defense (including the hyperparameters) the victim uses.
This is a common assumption used in the literature~\cite{DengYLZLXXW23Robust, SirinamIJW18, mathews2022sok}. 
The attacker is able to train using defended traces collected beforehand and tests on traces from the actual victim; this is known as adversarial training which we consider to be a realistic capability.

\section{Related Work}
\label{sec:related-work}


\subsection{Website Fingerprinting Attacks}
In general, a WF attack consists of three core parts: 
a meaningful trace representation of raw network traces, 
an effective model that takes the feature as input, 
and a training recipe that defines the training process and the loss function for model training. 
Existing WF attacks can be roughly classified into \textit{machine-learning-based (ML-based) attacks} and \textit{deep-learning-based (DL-based) attacks}, according to the models used in these attacks. 

\textbf{ML-based attacks.}
ML-based attacks generally involve extensive feature engineering to transform raw traces into meaningful feature vectors. 
These vectors are then utilized by specialized machine learning models for classification purposes. 
Representative techniques include SVMs~\cite{Cai12touching, WangG13improve, Panchenko16Web, PanchenkoNZE11}, kNN-based methods~\cite{WangCNJG14}, and Random Forests~\cite{Hayes16kfin}. 
The performance of ML-based attacks depends heavily on the quality of selected features. 
They are shown to be ineffective against recent defenses~\cite{shen2024real, gong2022surakav}.

\textbf{DL-based attacks.}
DL-based attacks automate the feature engineering process by using deep learning models to learn latent features. 
With packet sequences as input, these methods utilize various backbone architectures and train deep learning models in a supervised or semi-supervised manner.
These attacks are based on Stacked Denoising Autoencoder~\cite{abe2016fingerprinting}, Generative Adversarial Networks (GAN)~\cite{OhMRWH21}, Convolutional Neural Networks (CNN)~\cite{RimmerPJGJ18Automated, SirinamIJW18, Sirinam19Triplet, bhat2019var, mohammadTik19, shen23subverting, BahramaliBH23, DengHolmes24}, and transformers~\cite{JinLLS23Transformer, DengYLZLXXW23Robust, GuanXGLCL21}. 
As DL-based attacks show superior performance over ML-based ones on defended traces, we select the six most related and representative attacks to compare with our attack in this work. 

\begin{itemize} [left=10pt, itemsep=2pt]
    \item \textbf{DF}~\cite{SirinamIJW18}: DF attack designs a deep CNN model that takes the raw \textit{packet-direction} sequence as input. 
    The packet-direction sequence is a sequence of +1 and -1's to represent outgoing and incoming Tor cells. 
    (In Tor, each packet is a \textit{fixed-size cell} of 514 bytes.)

    \item \textbf{Tik-Tok}~\cite{mohammadTik19}: Tik-Tok uses the same model as DF, except that the input feature is the \textit{timing-with-direction} sequence (cell timestamps multiplied by +1 or -1 indicating their directions). 

    \item \textbf{Var-CNN}~\cite{bhat2019var}: Var-CNN applies a CNN model to predict both the direction sequence and the timing sequence and ensembles the results to get the final prediction.
    
    \item \textbf{RF}~\cite{shen23subverting}: RF is a new CNN-based attack using a new trace representation called TAM which aggregates the number of incoming and outgoing packets within fixed time windows of \SI{44}{ms}.
    It is robust against various defenses. 

    \item \textbf{ARES}~\cite{DengYLZLXXW23Robust}: ARES first uses a convolutional block to extract features on the packet ordering sequence, and then feeds the extract feature vector into a transformer for classification. 
    It is intended to attack multi-tab traces.
    We adapt it to attack single-tab traces. 

    \item \textbf{TMWF}~\cite{JinLLS23Transformer}: TMWF is another transformer-based attack for attacking multi-tab traces. 
    We adapt it to attack single-tab traces. 
\end{itemize}

\subsection{Website Fingerprinting Defenses}
As existing attacks are already highly accurate in fingerprinting undefended traces, we focus on developing a novel attack that breaks defended traces.
Existing defenses rely on adding extra non-informative (dummy) packets and delaying real packets to obscure the pattern of real traffic.
The core strategies of these defenses can be summarized as 
\textit{noise injection}~\cite{JuarezIPDW16, GongW20, AbusnainaJKNM20DFD, pulls20towards, LuoZCLCP11, CherubinHJ17}, 
\textit{traffic reshaping}~\cite{DyerCRS12, CaiNJ14, CaiNWJG14, LuBKD18, HollandH22, gong2022surakav}, 
\textit{pattern clustering}~\cite{NithyanandCJ14, WangCNJG14, WangG17, shen2024real}, 
\textit{adversarial perturbation}~\cite{li2022minipatch, shan2021patch, rahman21mockingbird, sadeghzadeh2021awa, LingXWGYF22, nasr21defeating, LiuDYSW22, jiang2024rudolf}, 
and \textit{traffic splitting}~\cite{HenriG0BT20, Cadena20Traffic}. 

\textbf{Noise injection.}
These defenses inject dummy data to obfuscate traffic.
WTF-PAD~\cite{JuarezIPDW16} injects noise during long cell intervals, while FRONT~\cite{GongW20} adds more cells at the beginning of the trace.
HTTPOS~\cite{LuoZCLCP11} and ALPaCA~\cite{CherubinHJ17} modify message sizes in the application layer by inserting dummy bytes.
DFD~\cite{AbusnainaJKNM20DFD} randomly injects dummy cells within each data burst.

\textbf{Traffic reshaping.} 
Defenses in this category alter traffic patterns by controlling the packet sending rate.
BuFLO Family members~\cite{DyerCRS12, CaiNJ14, CaiNWJG14, LuBKD18} standardize packet rates and pad the trace length.
Tamaraw~\cite{CaiNWJG14} fixes sending rates and extends traces to multiples of 100 cells.
RegulaTor~\cite{HollandH22} uses a decreasing cell sending rate over time, mimicking loading processes.
Surakav~\cite{gong2022surakav} generates traffic patterns with a trained generator, adjusting in real-time.

\textbf{Pattern clustering.}
Defenses cluster web pages into groups to produce uniform network patterns.
Supersequence~\cite{WangCNJG14} and Glove~\cite{NithyanandCJ14} calculate a super-trace for each cluster, ensuring identical traces within a group.
Palette~\cite{shen2024real} organizes pages into anonymization sets and dynamically refines the traffic pattern during loading.

\textbf{Adversarial perturbation.}  
Adversarial perturbations are meticulously crafted noise added to the input, designed to mislead deep learning models into making incorrect classifications.  
Numerous methods for crafting perturbations have been proposed in the context of images and videos~\cite{goodfellow15explaining, carlini17toward, liang2020efficient, liang2022parallel, liang2022large}.  
These methods provide new insights for designing defenses in website fingerprinting (WF) tasks.  
Such defenses aim to mislead deep learning models by introducing precisely crafted noise into the input data.  
Several studies propose specialized loss functions to minimize or maximize distances within the feature space~\cite{li2022minipatch, shan2021patch, rahman21mockingbird, sadeghzadeh2021awa, LingXWGYF22, nasr21defeating, LiuDYSW22, jiang2024rudolf}.  
However, these defenses face criticism for requiring access to the entire trace to compute the noise.  
Additionally, they are often ineffective against models that have been subjected to adversarial training~\cite{mathews2022sok}.  

\textbf{Traffic splitting.}
Defenses in this category aim to enhance security by routing traffic through multiple network paths, thereby preventing attackers on any single path from capturing enough packets to make accurate predictions. 
HyWF~\cite{HenriG0BT20} and TrafficSliver~\cite{Cadena20Traffic} use multihoming and multiple Tor circuits to distribute Tor cells.


\section{Attack Design}
\label{sec:attack-design}

In this section, we introduce our new attack \attack that can effectively undermine existing defenses.
\attack is short for \underline{W}ebsite \underline{F}ingerprinting with \underline{C}hannel-wise \underline{A}ttention on \underline{T}iming features which exploits timing features with an enhanced CNN-based backbone. 
We will first explain our intuition and then detail our design.

\subsection{Observation and Intuition} 

\begin{table*}[]
\centering
\caption{Comparison between trace representations used by different attacks. }
\label{tab:feature-representation-cmp}
\resizebox{0.95\textwidth}{!}{%
\begin{tabular}{@{}lccl@{}}
\toprule
\textbf{Trace representation} & \textbf{Description} & \textbf{Granularity$^*$} & \textbf{Representative Attacks} \\ \midrule
Statistical features & e.g., number of cells,  average loading times & $\fullmoon$ & kNN~\cite{WangCNJG14}, kFP~\cite{Hayes16kfin} \\
packet-direction sequence & a sequence of +1's and -1's & $\newmoon$ & AWF~\cite{RimmerPJGJ18Automated}, DF~\cite{SirinamIJW18} \\
timing-with-direction sequence & a sequence of timestamps multiplied by directions & $\newmoon$ & TikTok~\cite{mohammadTik19} \\
TAM & packet counts per time slot & $\LEFTcircle$ & RF~\cite{shen23subverting} \\
TAF & packet and burst size counts per time slot & $\LEFTcircle$ & Holmes~\cite{DengHolmes24} \\
\textbf{IAT histogram (ours)} & \textbf{inter-arrival-timing counts per time slot} & \textbf{$\LEFTcircle$} & \textbf{\attack} \\ \midrule
\multicolumn{4}{l}{$^*$ $\fullmoon$: coarse-grained granularity, $\LEFTcircle$: intermediate-grained granularity, $\newmoon$: fine-grained granularity}
\end{tabular}%
}
\end{table*}

A robust trace representation is crucial for the effectiveness of a WF attack. 
We explore the differences in trace representation among various attacks in Table~\ref{tab:feature-representation-cmp} and present the following observations:

\textbf{Statistical features are susceptible to manipulation by defenses.}
Most machine learning-based attacks rely on statistical features that are inadequate for comprehensive trace representation. 
These features can be deliberately manipulated to undermine the effectiveness of an attack, as demonstrated by the defense FRONT~\cite{GongW20}.
Given their susceptibility to defense strategies, we have opted not to utilize statistical features in our trace representation methodology.

\textbf{Timing features are useful in attacking defenses.}
Although Tik-Tok and DF share the same underlying architecture, Tik-Tok has been shown to surpass DF in overcoming certain defenses~\cite{mohammadTik19}, using a sequence of timing-with-direction as its input. 
This suggests that timing features are essential and should not be overlooked. 
Furthermore, many defenses incorporate time-sensitive mechanisms that activate at specific moments. 
For instance, RegulaTor dispatches all packets once the wait time for buffered data exceeds a set threshold~\cite{HollandH22}, 
and WTF-PAD adjusts its operational states based on the statistics of packet inter-arrival times~\cite{JuarezIPDW16}.
The activation timing of these mechanisms often correlates directly with the characteristics of the accessed web page. 
By leveraging timing information, these defenses’ vulnerabilities can be exploited. 
We incorporate inter-arrival times into our trace representation to undermine defenses. 

\textbf{Intermediate-grained features enhance robustness against defenses.}
Recent research indicates that using features with reduced granularity can significantly bolster the robustness of trace representations. 
A key example is TAM, a 2D matrix that logs the count of outgoing and incoming cells per time slot, as introduced by Shen et al.~\cite{shen23subverting}. 
Their findings reveal that two padding-based defenses, WTF-PAD and FRONT, disclose nearly as much information as in the undefended scenario when analyzed under TAM representation. 
This intermediate-grained approach mitigates the effects of localized perturbations, enabling the model to more accurately capture the overarching trends of a trace’s loading process. 
TAF, a variant of TAM, further incorporates two additional aggregated features related to burst sizes per time slot~\cite{DengHolmes24}.  
However, both TAM and TAF do not fully harness the timing details of individual packets within each time slot. 
In contrast, our trace representation, which also adopts an intermediate granularity, enriches this aspect by incorporating timing data, thereby enhancing the performance of our attack (See Section~\ref{sec:a-new-trace-representation}).

\subsection{A New Trace Representation}
\label{sec:a-new-trace-representation}

Based on our observations, we propose a novel trace representation, the \textbf{Inter-Arrival Time (IAT) histogram}, 
which enhances the learning capacity of deep learning models for web page analysis. 
Our approach constructs an intermediate-grained feature representation that captures inter-arrival timing information within fixed time slots.

Before detailing our algorithm for computing IAT histograms, we first formally define \textbf{traces} in the Tor network. 
A trace is an ordered sequence of \(N\) Tor cells, sorted by their timestamps:
\setlength{\abovedisplayskip}{4pt}
\setlength{\belowdisplayskip}{4pt}
\begin{equation}
   X = (p_0, p_1, \dots, p_{N-1}).
\end{equation}
Each Tor cell is represented by a timestamp and a direction:
\begin{equation}
p_i = (t_i, d_i),
\end{equation}
where \(t_i \geq 0\) denotes the timestamp, and \(d_i \in \{+1, -1\}\) represents the direction of the cell. Since each Tor cell is encrypted and has a fixed size, we do not record the bytes. By convention, \(d_i = 1\) indicates an outgoing cell.
The inter-arrival time of the $i$-th cell $p_i$ is defined as 
\begin{equation}
\delta_i = 
\begin{cases} 
0, & \text{if } i = 0, \\
t_i - t_{i-1}, & \text{if } 0 < i \leq N. 
\end{cases}
\end{equation}

\begin{figure}
    \centering
    \includegraphics[width=\linewidth]{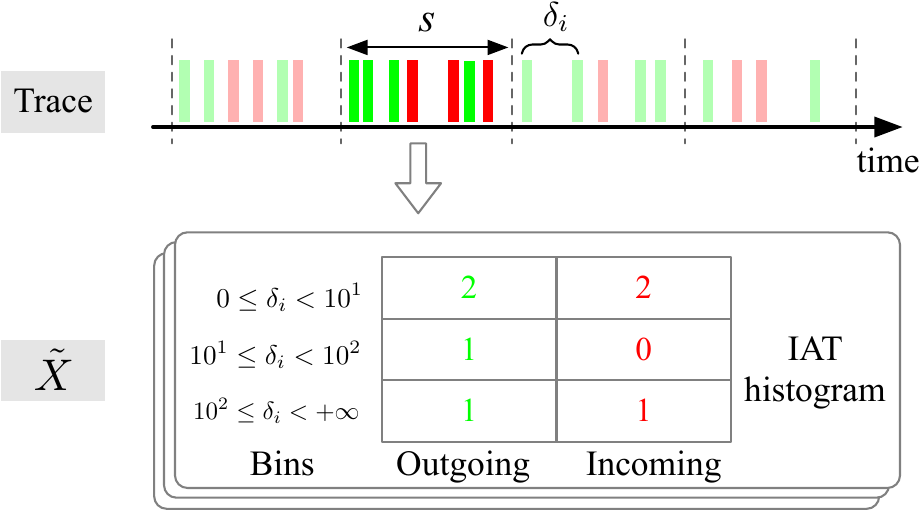}
    \caption{Visualization of IAT histogram computation. In this example, we bin all the cells according to their IAT values (in milliseconds) into $G=3$ distinct bins.}
    \label{fig:iat-computation}
    \vspace{-7pt}
\end{figure}

Given a trace $X$, we calculate the trace representation $\Tilde{X}$ as shown in Figure~\ref{fig:iat-computation}. 
We first divide the trace into fixed time slots of duration $s$.
($s$ is a hyperparameter.)
For each time slot, we compute a distribution of IAT values for both incoming and outgoing cells within it.
For convention, we define the set of all cells within the $k$-th time slot as $\mathcal{S}_k = \{p_i \mid k \cdot s \leq t_i < (k+1) \cdot s, 0 \leq i < N\}$ for $0 \leq k < L$. 
Furthermore, we denote the set of all outgoing cells in $\mathcal{S}_k$ as $\mathcal{S}_k^+$ and the set of all incoming cells as $\mathcal{S}_k^-$. 
We compute two histograms that represent the distributions of IAT values for the outgoing and incoming cells in $\mathcal{S}_k$, respectively. 
Specifically, we count the number of cells that fall into each IAT bin: 
\begin{align} 
    \Tilde{X}[r, 0, k] & = |\{ p_i \mid b_r \leq \delta_i \leq b_{r+1} \} \cap \mathcal{S}_k^+|, \\
    \Tilde{X}[r, 1, k] & = |\{ p_i \mid b_r \leq \delta_i \leq b_{r+1} \} \cap \mathcal{S}_k^-|, 
\end{align} 
where $|\cdot|$ returns the cardinality of the set (i.e., the number of cells), and $\mathcal{B} = \{b_r \in \mathbb{R} \mid r = 0, 1, \cdots, G\}$ represents the boundary values that evenly divide all IAT values on a logarithmic scale ($G$ bins in total). 
We set $b_0 = 0$ and $b_{G} = +\infty$. 
After the computation for $L$ time slots, we get a matrix of shape $G \times 2 \times L$ as our feature representation.
The hyperparameter $L$ is the total number of time slots considered for a trace, while the hyperparameter $G$ is the number of bins to gather the IAT values in each time slot.

We chose logarithmic bins for constructing the inter-arrival time (IAT) histogram because the distribution of inter-arrival times in network traffic is often highly skewed. 
Many inter-arrival times are concentrated in shorter durations, with fewer instances spanning longer durations. 
Logarithmic bins are better suited to capture this variability because they provide finer granularity for smaller values, where more data points are concentrated, while still representing larger values efficiently.

The IAT histogram not only captures the volume of data but also the timing information; 
it reflects changes in the \textit{density} of cells over time.
IAT values can reveal dependency relationships between different cells.
Given that different web pages typically have varying layouts and resource counts, web servers may respond at different rates, resulting in distinctive IAT values. 
Instead of using precise IAT values, we opt to bin them into a few intervals to achieve a more robust feature representation. Tor circuit latency applies an inherently random multiplicative effect on packet timing, so we use exponential bins.
However, not all IAT intervals prove informative; 
thus, we introduce a new CNN block designed to automatically learn the significance of different IAT intervals, enhancing our utilization of the trace representation (detailed in Section~\ref{sec:model-architecture}).

\subsection{Model Architecture}
\label{sec:model-architecture}

\begin{figure}
    \centering
    \includegraphics[width=0.95\linewidth]{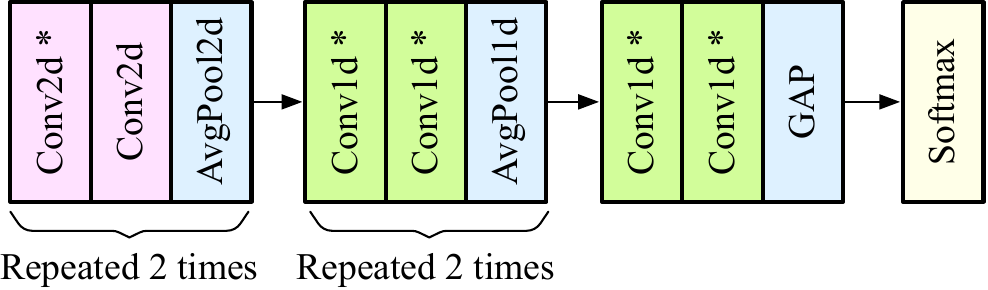}
    \caption{\attack model's architecture. *The \textit{first} Conv2d and all Conv1d blocks are our proposed new blocks.}
    \label{fig:architecture}
\end{figure}

We design a new CNN-based model for our proposed attack \attack that can effectively handle our trace representation.
In general, the IAT histogram first passes through a 2D convolutional module, followed by a 1D convolutional module, and finally a global average pooling (GAP) module, as shown in Figure~\ref{fig:architecture}.

The 2D convolutional module is responsible for extracting local features from the IAT histogram $\Tilde{X}$.
The input is processed through two Conv2d blocks and an average pooling layer.
This sequence is repeated twice before being passed into the 1D convolutional module.
A dropout layer is added after the average pooling layer to mitigate overfitting.
Except for the first Conv2d block, each Conv2d block consists of a 2D convolutional layer, a batch normalization layer~\cite{BN}, and a GELU activation layer~\cite{gelu}.

The 1D convolutional module is tasked with extracting higher-level features after all local features have been gathered.
Following the RF design~\cite{shen23subverting}, we utilize global average pooling (GAP) to convert the hidden features into $C$ logits for final prediction.
Here, $C$ denotes the number of classes.
It has shown that GAP is more effective at preventing overfitting compared to using fully connected layers to produce logits~\cite{shen23subverting}.

The key components of our model that are different from other attacks are detailed as follows.

\begin{figure}
    \centering
    \includegraphics[width=\linewidth]{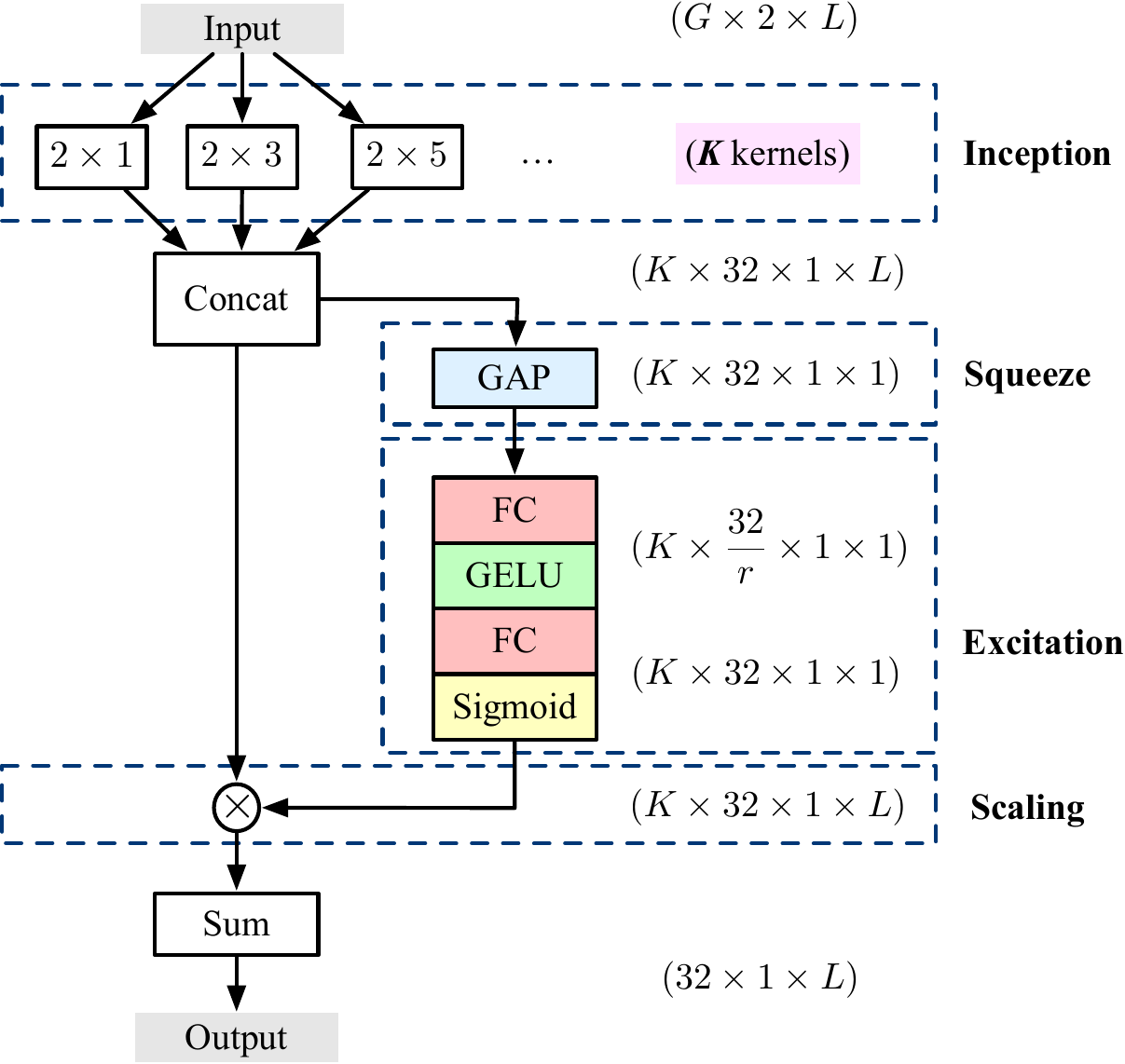}
    \caption{Illustration of the first Conv2d block: utilizing multiple kernels for feature extraction at various scales via an Inception block, followed by fusion of features across different channels with learned weights in the SEBlock.}
    \label{fig:first-block}
\end{figure}

\textbf{Inception2d with SEBlock.}
We introduce this innovative block as the first component of our model to enhance feature capture within $\Tilde{X}$.
As depicted in Figure~\ref{fig:first-block}, we initially apply an Inception block~\cite{SzegedyLJSRAEVR15} that utilizes $K$ kernels to extract features from $\Tilde{X}$.
The kernel width is consistently set to 2 to capture the spatial correlation between incoming and outgoing cells.
The kernel height is defined as $2k+1$ $(k=0,1,\cdots, K-1)$, which improves the model's ability to discern features at varying scales.
Subsequently, a Squeeze-and-Excitation Block (SEBlock)~\cite{HuSS18} is employed to ascertain the channel-wise importance of the features, which are then multiplied by the output from the Inception block.
The SEBlock, a prominent component in CNN architectures, is designed to enhance network representational power by prioritizing informative features and suppressing less useful ones.
This involves a \textbf{Squeeze} step that compresses each channel to a single numerical value through global average pooling across the spatial dimensions of the feature map,
followed by an \textbf{Excitation} step that processes these descriptors through a two-layer fully connected (FC) network, initially reducing the dimensionality (channels: $32 \rightarrow 32/r$) and subsequently restoring it.
For our model, we set $r=16$.
The final output is a set of scaling weights between 0 and 1, achieved through a Sigmoid activation.

This block significantly improves our model's ability to interpret trace representations from multiple scales and accentuate the salient features across different channels.
To minimize computational demands, this block is only applied at the input stage of the model.

\textbf{Inception1d Block.}
The output from the 2D convolutional module is reshaped into a 1D feature map before being introduced to the 1D convolutional module.
Each Conv1d block illustrated in Figure~\ref{fig:architecture} constitutes an Inception1d Block, which is followed by a batch normalization layer and a GELU activation layer.
Similar to the Inception2d Block, we deploy several kernels of size $2k+1$ ($k=0,1,\cdots, K-1$) to extract and fuse features together.
At this stage, we do not apply the SEBlock as it offers marginal benefits while significantly increasing computational costs.

\section{Attack Evaluation}
\label{sec:attack-evaluation}
In this section, we evaluate the performance of \attack.
We begin by describing the experimental setup and datasets used in our study.
Next, we present the hyperparameter tuning process for our attack.
Following this, we conduct extensive experiments in both closed-world and open-world scenarios to compare \attack with state-of-the-art (SOTA) attacks.
Finally, we perform an ablation study to analyze the impact of each component of \attack.

\subsection{Experiment Setup}
\label{sec:experiment-setup}

We make use of the WFDefProxy framework~\cite{GongZZW24}, which is specifically designed for collecting live Tor traces.
We have rent two servers from Google Cloud, designating one as our Tor client and the other as the Tor entry node. 
The client, located in Singapore, operates on Ubuntu 20.04 LTS and is equipped with 4 CPU cores and 16 GB of memory.
The Tor entry node is strategically placed in America to ensure a physical distance from the client, running Debian 10.6 with the Tor binary version 0.4.4.5.
To optimize the collection process, we dockerize the Tor client, enabling the parallel operation of multiple clients to reduce the collection time.
The version of the Tor browser used is 12.0.

\textbf{Datasets.}
To evaluate the attacks, we initially collected a new undefended dataset using WFDefProxy.
This dataset comprises Tor traces from 100 monitored pages and 10,000 non-monitored pages.
Each monitored page was loaded 100 times, while each non-monitored page was loaded once.
We discarded instances from inaccessible pages or unsuccessful loadings, specifically those where the cell count was less than 50, during our collection process.
The monitored pages were selected from the top 100 sites on the Tranco list~\cite{LePochat2019}, while the non-monitored pages were taken from those ranked 200th and beyond in the same list.
The Tranco list is a ranking of websites based on their popularity, designed specifically for research purposes. 
It aggregates data from multiple sources over a period of time, making it more stable and less prone to manipulation.

Due to the absence of accurate simulation code for Surakav~\cite{gong2022surakav}, we collected another dataset defended by Surakav using Gong's implementation on WFDefProxy.
We adhered to the same methodology employed in the collection of the undefended dataset.
The total data collection spanned one month.

\textbf{Evaluated attacks and defenses.}
We compare our attack with four CNN-based attacks: DF~\cite{SirinamIJW18}, Tik-Tok~\cite{mohammadTik19}, Var-CNN~\cite{bhat2019var}, and RF~\cite{shen23subverting}; 
and two Transformer-based attacks, ARES~\cite{DengYLZLXXW23Robust} and TMWF~\cite{JinLLS23Transformer}.
Following the recommended hyperparameters, we tune these attacks on our datasets to achieve their optimal performance.
All the attacks are trained using an H100 card with 47~GB of memory.

We evaluate our attack against a variety of defenses, 
including two noise-injection-based defenses, WTF-PAD~\cite{JuarezIPDW16} and FRONT~\cite{GongW20}; three traffic-reshaping-based defenses, RegulaTor~\cite{HollandH22}, Tamaraw~\cite{CaiNWJG14}, and Surakav~\cite{gong2022surakav}; 
one traffic-splitting-based defense, TrafficSliver~\cite{Cadena20Traffic};
and one pattern-clustering-based defense, Palette~\cite{shen2024real}.
We use simulation code provided by the authors to generate defended datasets based on our undefended dataset.
Since Surakav does not have accurate simulation, we directly collect a Surakav dataset in the real Tor network.

\textbf{Evaluation methodology.}
Following the evaluation methodology of prior work~\cite{gong2022surakav, GongW20}, we divide the dataset into training, validation, and test sets using an 8:1:1 ratio.
We fine-tune the attack model using the validation set.
To ensure robustness and mitigate bias, we conduct 10-fold cross-validation for each experiment and report the combined results.

We use \textit{classification accuracy} to assess the attack's performance in a closed-world scenario.
Accuracy is calculated as the ratio of correctly predicted instances to the total number of test instances.
In the open-world scenario, we evaluate performance using the \textit{Precision-Recall Curve} to explore the precision-recall trade-off across various confidence thresholds of the model.

Following prior work~\cite{Wang20HPOWF}, we compute precision and recall as follows.
We denote a trace $X$ where its ground truth label is $y^{true} \in \{0, 1, \cdots, C-1, C\}$, with $y^{true} = C$ indicating a non-monitored trace.
The output of an attack model is denoted as $\bm{y}$, a $(C+1)$-dimensional vector where each element reflects the confidence value of being predicted as that class.
The index with the highest confidence value is denoted as $j^* = \arg\max_{0 \leq j \leq C} y_j$, and the maximum confidence value is $y_{j^*}$.
Given a confidence threshold $\tau$, a trace is predicted as monitored if and only if $j^* < C$ and $y_{j^*} > \tau$.
A True Positive (TP) occurs when a monitored trace is correctly predicted in its class.
A Wrong Positive (WP) occurs when a monitored trace is predicted in another incorrect monitored class.
A False Negative (FN) happens when a monitored trace is predicted to be non-monitored.
A False Positive (FP) occurs when a non-monitored trace is predicted to be a monitored class.
Precision is computed as TP/(TP+WP+FP), and recall is computed as TP/(TP+WP+FN).
By varying the confidence threshold $\tau$ from 0 to 1, we derive multiple tuples of (precision, recall) values and plot the Precision-Recall Curve.
We also use \textit{F1-score} to comprehensively evaluate an attack's performance, calculated as $2 \times Precision \times Recall / (Precision + Recall)$. 

\textbf{Ethical consideration.} 
We adhered to prior work to carefully mitigate ethical concerns in our data collection process.
First, we utilized Python scripts to automate the Tor browser, ensuring that none of the collected traffic originated from actual users. 
Second, we retained only the essential information for each trace (i.e., cell directions and timestamps). 
Lastly, we limited the number of parallel clients to five during the crawling process, thereby minimizing the potential impact on the Tor network.

\subsection{Hyperparameter Tuning of \attack}

\begin{table}[]
\centering
\caption{Hyperparameter tuning for \attack model}
\label{tab:parameter-tuning}
\resizebox{\columnwidth}{!}{%
\begin{tabular}{@{}ccc@{}}
\toprule
\textbf{Hyperparameter}     & \textbf{Search Space} & \textbf{Final} \\ \midrule
Trace Length $L$            & [500, ..., 3000]      & 1800           \\
Time Slot $s$ (ms)          & [22, ..., 330]        & 44             \\
Bin Number $G$              & [2, ..., 10]          & 9              \\
Inception Kernel Number $K$ & [2, ..., 9]           & 4              \\
Optimizer                   & [Adam, Adamax, SGD]   & Adam           \\
Learning rate               & [1e-5, ..., 5e-3]     & 1e-3           \\
Weight Decay                & [1e-5, ..., 1e-3]     & 5e-4           \\
Batch Size                  & [64, 128, 256]        & 64             \\
Epoch Number                & [20, ... , 80]        & 50             \\ \bottomrule
\end{tabular}%
}
\end{table}

We prototype our attack using Pytorch~2.3.1 with 2,000 lines of code. 
We use the undefended dataset to tune our attack, observing the closed-world accuracy on the validation set. 
The hyperparameter search space and the optimal values are summarized in Table~\ref{tab:parameter-tuning}. 
We used the ASHA scheduler~\cite{Li2020Tuning}, a state-of-the-art hyperparameter tuning algorithm, and sampled 2,000 points within the search space. 

The trace length $L$ and the time slot $s$ determine which cells are considered in a trace.
We have determined that setting $L=1,800$ and $s = \SI{44}{ms}$ (i.e., considering cells in the first \SI{80}{s}) yields the best performance.
The parameter $G$ is the number of bins used to group all IAT values.
Increasing the number of bins enhances granularity but also adds computational overhead.
We have found that $G=4$ provides optimal performance.
The number of inception kernels is crucial for extracting features from various scales.
Using more kernels allows for the fusion of richer feature information, although it increases computational costs.
We have found that $K=4$ in both the 1D and 2D CNN modules yields the best performance.

\subsection{Closed-world Evaluation}
\label{sec:closed-world-eval}

\begin{table*}[]
\centering
\caption{Attack accuracy (\%) against different defenses in the closed-world scenario.
\attack demonstrates the highest performance against 6 out of 7 defenses (bolded).
The accuracy differences between \attack and the second-best attack are highlighted, with red indicating higher results.}
\label{tab:closed-world-evaluation}
\resizebox{0.95\textwidth}{!}{%
\begin{tabular}{@{}cc|cc|ccccccc@{}}
\toprule
\multirow{2}{*}{\textbf{Type}} & \multirow{2}{*}{\textbf{Defense}} & \multicolumn{2}{c|}{\textbf{Overhead}} & \multirow{2}{*}{\textbf{TMWF}} & \multirow{2}{*}{\textbf{ARES}} & \multirow{2}{*}{\textbf{VarCNN}} & \multirow{2}{*}{\textbf{DF}} & \multirow{2}{*}{\textbf{TikTok}} & \multirow{2}{*}{\textbf{RF}} & \multirow{2}{*}{\textbf{\attack (ours)}} \\
 &  & DO & TO &  &  &  &  &  &  &  \\ \midrule
 & Undefended & 0 & 0 & 75.96 & 91.14 & 91.35 & 93.44 & 93.41 & 92.39 & \textbf{94.47}\uptrend{1.03} \\
\multirow{2}{*}{\textbf{\begin{tabular}[c]{@{}c@{}}Noise\\ Injection\end{tabular}}} & WTF-PAD & 23 & 0 & 73.58 & 84.26 & 78.49 & 86.28 & 86.57 & 87.88 & \textbf{93.50}\uptrend{5.62} \\
 & FRONT & 76 & 0 & 25.93 & 55.42 & 45.07 & 48.64 & 49.26 & 85.24 & \textbf{93.18}\uptrend{7.94} \\
\multirow{3}{*}{\textbf{\begin{tabular}[c]{@{}c@{}}Traffic\\ Reshaping\end{tabular}}} & RegulaTor & 45 & 23 & 9.64 & 13.67 & 11.52 & 12.98 & 24.70 & 38.48 & \textbf{47.78}\uptrend{9.30} \\
 & Surakav & 103 & 23 & 13.33 & 11.15 & 7.79 & 12.26 & 15.04 & 30.92 & \textbf{59.12}\uptrend{28.20} \\
 & Tamaraw & 173 & 34 & 8.88 & 10.38 & 10.85 & \textbf{11.07} & \textbf{11.07} & 8.87 & 8.04\downtrend{3.03} \\
\textbf{Clustering} & Palette & 131 & 6 & 10.39 & 6.63 & 5.42 & 5.17 & 6.24 & 15.51 & \textbf{16.48}\uptrend{0.97} \\
\textbf{Splitting} & TrafficSliver & 0 & 0 & 5.78 & 7.74 & 15.12 & 5.64 & 14.67 & 39.88 & \textbf{50.12}\uptrend{10.24} \\ \bottomrule
\end{tabular}%
}
\end{table*}

In this section, we evaluate the performance of \attack against the SOTA defenses in the closed-world scenario where the attacker tries to distinguish 100 monitored classes between each other. 
We compare our attack with the other six attacks as described in Section~\ref{sec:experiment-setup}. 

As shown in Table~\ref{tab:closed-world-evaluation}, we evaluate our approach against seven distinct defenses, categorized by their types of mechanisms. 
In the table, we also report the overheads associated with each defense. 
The data overhead is measured as the ratio of the number of dummy cells to the number of real cells across the entire dataset; data overhead burdens the network. 
Similarly, the time overhead is calculated as the ratio of the additional time required to load a page to the original loading time, measured across the entire dataset; time overhead affects user experience. 
Overall, \attack demonstrates superior attack performance, outperforming existing methods on both the undefended dataset and six out of the seven defended datasets.

\textbf{Noise-injection-based defenses are significantly undermined by \attack}.
For instance, WTF-PAD proves ineffective against most attacks.
Similarly, although FRONT incurs a substantial 76\% data overhead on our dataset, it remains largely ineffective; 
our attack results in only a 1\% drop in accuracy ($94\% \rightarrow 93\%$).
In contrast, the second-best attack, RF, achieves 85\% accuracy against FRONT.
Comparatively, other attacks that utilize fine-grained features suffer approximately a 50\% accuracy loss when tested against FRONT.
These results demonstrate the robustness of semi-grained features against noise-injection-based defenses.

\textbf{Timing-sensitive defenses can be greatly exploited by \attack}. 
Strong defenses typically employ mechanisms designed to strategically delay cells at specific times, which are triggered by the characteristics inherent to the page. 
Such mechanisms often inadvertently leak information, a vulnerability confirmed by our experiments. 
For example, in our tests, TikTok and DF, which differ mainly in their use of timing information at the input stage, show distinct performance outcomes. 
TikTok achieves a 12\% higher accuracy against RegulaTor (25\% vs. 13\%) and a 3\% higher accuracy against Surakav (15\% vs. 12\%) compared to DF. 
Additionally, \attack capitalizes on this timing information to achieve an accuracy of 48\% against RegulaTor. 
Surakav, once considered the SOTA defense for this category, is compromised by \attack, which achieves a notably high accuracy of 59\%. 
In contrast, the previously best-performing attack, RF, manages only a 31\% accuracy against Surakav. 
The effectiveness of \attack stems from its adept utilization of timing information in both our trace representation and the model's backbone. 

However, our defense and other attacks have not compromised Tamaraw, which is provably secure. 
Tamaraw sends cells at constant rates, irrespective of the page being loaded, thereby leaking minimal timing information. 
Despite its effectiveness, the high overhead incurred by Tamaraw hinders its practical deployment.

\textbf{Splitting-based defenses are not as secure as expected}.
Splitting-based defenses, proposed as a novel solution against website fingerprinting, are touted for their minimal overhead and robust security level. 
The Tor project has officially proposed the implementation of traffic splitting~\cite{torproposal329}. 
Despite these advantages, our results indicate that these defenses remain vulnerable under \attack. 
\attack achieves an accuracy of over 50\% against TrafficSliver, significantly outperforming other methods like RF (40\%) and VarCNN (15\%). 
All other attacks tested register accuracies below 15\%.

In summary, our experiments demonstrate that \attack significantly outperforms all other attacks in the closed-world scenario. 
Particularly effective against timing-sensitive defenses, \attack leverages timing information within its design to achieve superior results.

\subsection{Open-world Evaluation}
\label{sec:open-world-eval}

\begin{figure}[t]
    \centering
    \includegraphics[width=\linewidth]{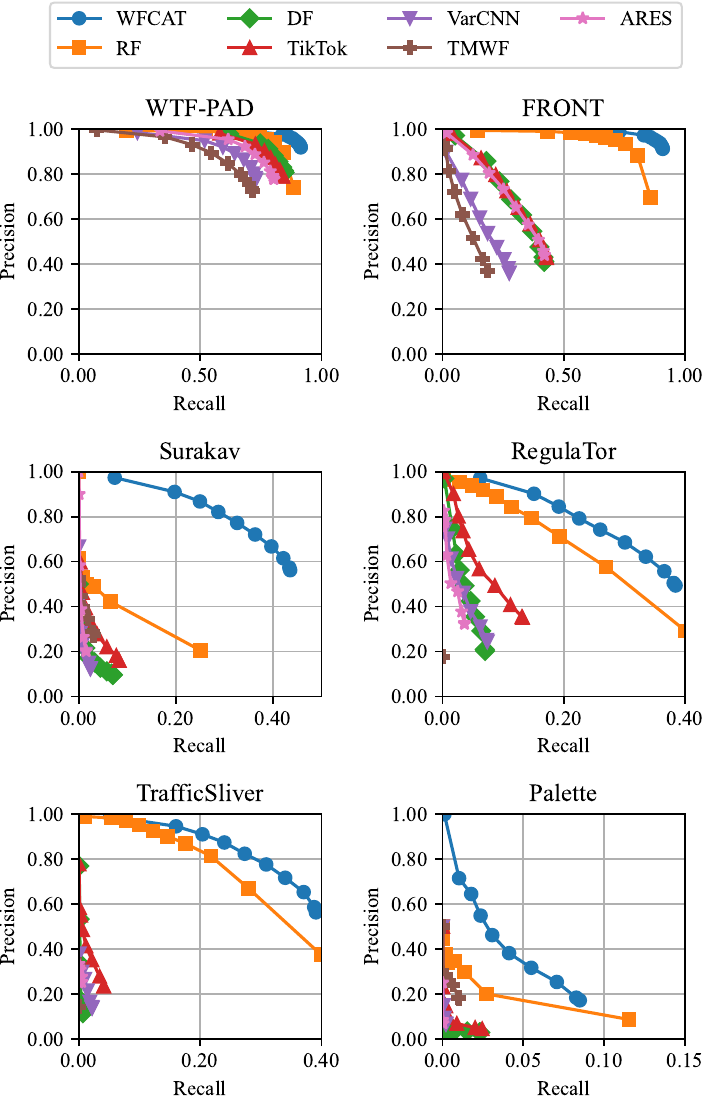}
    \caption{Attack performance against various defenses in the open-world scenario. The \attack method significantly outperforms the other attacks against all defenses. Note that the x-axis scale may vary across subfigures.}
    \label{fig:open-world-prcurve}
\end{figure}

Next, we evaluate the performance of \attack in the more realistic open-world scenario where the attacker tries to figure out whether or not the victim is visiting a specific monitored or a non-monitored page. 

\textbf{Performance on the undefended dataset.}
Most attacks exceed 0.90 in both precision and recall; our attack leads with a 0.93 F1-score.
On the Tamaraw dataset, despite all defenses showing low precision and recall, the highest performance (0.04 F1-score) is seen in DF and TikTok, with our attack at 0.02 F1-score.
Tamaraw, despite its effectiveness against all attacks due to uniform traffic formatting, proves too costly for practical use.

\textbf{Performance on the defended datasets.}
Figure~\ref{fig:open-world-prcurve} compares the performance of our attack with other attacks on different defended datasets.
In general, \attack outperforms all the other six attacks against all the defenses in the open-world scenario.
The second best attack is RF.
Compared to the other two timing-based attacks, TikTok and VarCNN, we find that \attack achieves significantly better results.
Both \attack and RF use intermediate-grained feature as input, which indicates that such a feature representation form is more robust against noise.

Similar to the closed-world scenario, the noise-injection-based defenses, WTF-PAD and FRONT, show little impact on \attack's performance in the open-world scenario.
It achieves a 0.92 F1-score against both FRONT and WTF-PAD (0.01 drop compared to the undefended case).
This confirms that simply injecting random noise into the trace is not effective against our attack.

The most surprising result is on Surakav, which was considered to be the best defense using traffic reshaping techniques.
\attack has partially broken it, achieving 0.56 precision and 0.44 recall.
It outperforms other attacks by a large margin. 
For example, RF only achieves 0.20 precision and 0.25 recall.
The other attacks fail to attack Surakav (F1-score below 0.10).
The key reason why Surakav is vulnerable to \attack is that Surakav's padding mechanism is time-sensitive, which can be captured and learnt by \attack.
This is also true for RegulaTor.

TrafficSliver was previously a highly effective defense with nearly zero overhead.
It was weakened by RF, achieving a recall of 0.37 and a precision of 0.40.
\attack successfully improves the recall to over 0.59 while maintaining a similar precision of 0.39.

\subsection{Impact of Circuit Bandwidth}

\begin{table*}
\centering
\caption{Attack Accuracy (\%) with training traces from fast (resp. slow) circuits and testing traces from slow (resp. fast) circuits. The best performance on each dataset is marked in bold font. \attack significantly outperforms the other attacks in most cases.}
\label{tab:slow-fast-circuit}
\resizebox{\textwidth}{!}{%
\begin{tabular}{@{}clcccccc|cccccc@{}}
\toprule
\multicolumn{2}{c}{\textbf{}} & \multicolumn{6}{c|}{\textbf{Train on slowest traces, test on fastest traces}} & \multicolumn{6}{c}{\textbf{Train on fastest traces, test on slowest traces}} \\ \cmidrule(l){3-14} 
\multicolumn{2}{c}{\textbf{}} & WTF-PAD & FRONT & RegulaTor & Surakav & Palette & TrafficSliver & WTF-PAD & FRONT & RegulaTor & Surakav & Palette & TrafficSliver \\ \midrule
\multicolumn{2}{c}{\textbf{TMWF}} & 80.4 & 21.8 & 6.8 & 7.8 & 7.3 & 5.1 & 47.8 & 11.5 & 4.2 & 10.3 & 6.0 & 5.5 \\
\multicolumn{2}{c}{\textbf{ARES}} & 83.7 & 42.7 & 10.5 & 5.8 & 4.9 & 7.4 & 49.3 & 40.1 & 11.3 & 7.5 & 5.0 & 7.6 \\
\multicolumn{2}{c}{\textbf{VarCNN}} & 79.1 & 39.0 & 9.8 & 3.6 & 3.6 & 7.5 & 46.2 & 28.7 & 5.1 & 5.9 & 5.6 & 3.5 \\
\multicolumn{2}{c}{\textbf{DF}} & 86.7 & 38.2 & 10.7 & 7.1 & 3.2 & 7.0 & \textbf{59.4} & 30.7 & 11.9 & 8.6 & 4.2 & 5.7 \\
\multicolumn{2}{c}{\textbf{TikTok}} & 87.0 & 34.5 & 13.7 & 7.9 & 3.2 & 7.0 & 58.9 & 30.4 & 14.1 & 9.5 & 6.4 & 6.7 \\
\multicolumn{2}{c}{\textbf{RF}} & 90.9 & 91.7 & \textbf{22.7} & 32.1 & 8.1 & 42.4 & 46.1 & 48.2 & 19.7 & 15.8 & 9.2 & 14.3 \\
\multicolumn{2}{c}{\textbf{\attack}} & \textbf{93.4} & \textbf{95.6} & 20.9 & \textbf{48.0} & \textbf{8.5} & \textbf{48.8} & 56.8 & \textbf{59.4} & \textbf{27.3} & \textbf{32.1} & \textbf{9.5} & \textbf{20.1} \\ \bottomrule
\end{tabular}%
}
\end{table*}

In practice, victims traverse circuits with varying latency and bandwidth, which results in differing page load times. 
Prior research has demonstrated that discrepancies in network conditions between training and testing traces can significantly degrade attack performance~\cite{BahramaliBH23}. 
To explore this issue, we split a defended dataset based on trace load times. 
Each attack is trained on the 80\% fastest (resp. slowest) traces per page and tested on the 10\% slowest (resp. fastest) traces, with the remaining 10\% reserved for validation. 
We present the closed-world accuracy for each attack in Table~\ref{tab:slow-fast-circuit}.
The findings are as follows. 

\textbf{Poor network conditions adversely affect WF attacks.}
We observed that traces from the fastest circuits are significantly easier to fingerprint by all attacks. 
This is primarily because poor network conditions can lead to partial page loads, packet drops, or retransmissions, thereby making the traces noisier for the attacker. 
While a few studies explore enhancing attack performance through data augmentation and contrastive learning~\cite{BahramaliBH23, Sirinam19Triplet, XieContrastive24}, these approaches are orthogonal to our work. 
Our focus is on comparing the raw performance of these attacks without any additional training techniques.

\textbf{\attack shows more robust performance than other attacks.}
Across all tests, \attack achieves the highest accuracy in attacking five out of six defenses for both the fastest and slowest traces. 
While testing on the fastest traces, \attack exhibits minimal accuracy loss in attacks against defenses like WTF-PAD, FRONT, and TrafficSliver, compared to the results in Table~\ref{tab:closed-world-evaluation}. 
However, time-sensitive defenses such as RegulaTor and Surakav see some impact on performance due to bandwidth mismatches. 
Despite the challenges posed by testing on the slowest traces, \attack remains the strongest performer among all considered attacks. 
Notably, \attack outperforms the second-best attack by margins of 16\% on Surakav (32\% vs. 16\%), 11\% on FRONT (59\% vs. 48\%), and 7\% on RegulaTor (27\% vs. 20\%).

\subsection{Impact of Training Sample Number}

\begin{figure}
    \centering
    \includegraphics[width=\linewidth]{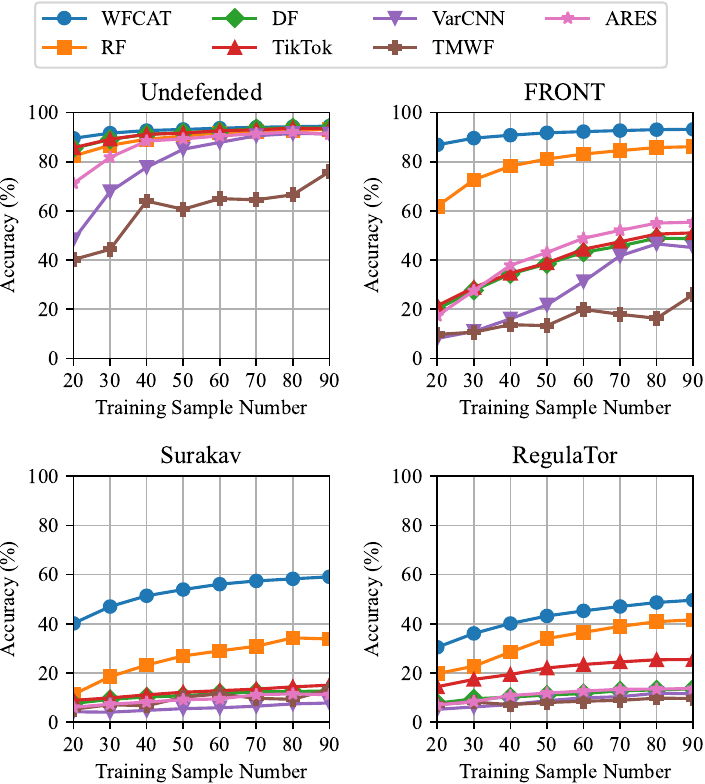}
    \caption{Attack performance with different number of training samples in the closed-world scenario. \attack requires fewer samples to achieve an even better performance on four datasets.}
    \label{fig:training-sample-num}
\end{figure}

As the data collection process is inherently time-consuming, we are particularly interested in how many training samples each attack requires to achieve satisfactory performance. 
Moreover, given that most web pages are constantly evolving, an attack's ability to rapidly adapt with fewer training samples is crucial for maintaining robust performance. 
We assessed the performance of our attack across four datasets: the undefended dataset, which serves as the baseline, and three datasets with defenses -- FRONT, Surakav, and RegulaTor. 
We vary the number of training samples per class from 20 to 90 and observe the attack accuracy in a closed-world scenario. 
Figure~\ref{fig:training-sample-num} shows the results.

We discovered that \textbf{\attack requires fewer samples than other attacks to converge to optimal performance}. 
For instance, on the undefended dataset, with only 30 samples per class, our attack achieves a remarkable 92\% accuracy. 
In contrast, RF and TikTok require 60 samples, and ARES needs 80 samples to reach similar outcomes. 
Notably, VarCNN and TMWF do not reach a plateau even with 90 samples. 
The superiority of \attack is further underscored in defended datasets. 
For the FRONT dataset, with merely 30 samples per class, \attack surpasses all other six attacks that train with 90 samples per class (i.e., the full training set). 
For Surakav, \attack trained on 20 samples per class achieves 6\% higher accuracy than the next best attack, RF, trained on the full set. 
For RegulaTor, \attack consistently maintains around 10\% higher accuracy than RF with different number of training samples.

To conclude, our attack requires fewer training samples to effectively compromise the state-of-the-art defenses compared to other attacks.

\subsection{Comparison of Training Time}

\begin{figure}
    \centering
    \includegraphics[width=0.95\linewidth]{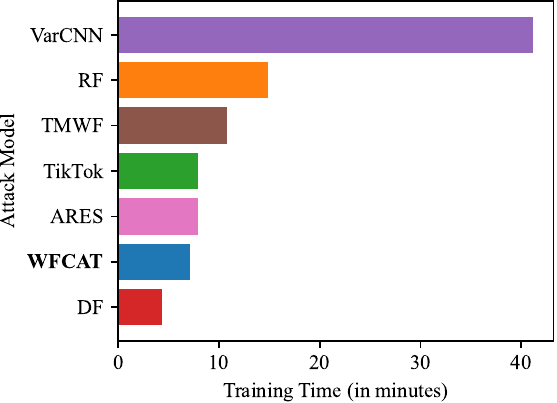}
    \caption{Training time of attacks on the open-world undefended dataset. \attack has the second shortest training time.}
    \label{fig:training-time}
\end{figure}

Apart from the number of training samples, training time is another significant cost that attackers must consider. 
In this section, we compare the training time of our attack with that of other attacks using the undefended dataset we collected. 
We calculated the average training time across 10 cross-validation folds, and the results are presented in Figure~\ref{fig:training-time}.

Among all seven attacks, our attack ranks as the second fastest. 
It requires only 7 minutes to complete training, slightly longer than the fastest attack, DF, which completes in 4 minutes. 
This minor increase in time is inevitable due to the additional blocks introduced into the network to enhance performance. 
However, we have minimized overhead by eliminating unnecessary complexity in our design. 
Notably, the previously best-performing attack in terms of accuracy, RF, requires nearly 15 minutes for training, which is more than double the time required by \attack. 
TikTok, ARES, and TMWF all require slightly more training time than \attack, while VarCNN takes the longest at nearly 42 minutes. 
This extended duration is due to its ensemble mechanism, where timing and directional sequences are processed through two sub-networks. 
Despite this, it does not yield effective performance, as discussed in Sections~\ref{sec:closed-world-eval} and \ref{sec:open-world-eval}, highlighting its inefficient use of time information. 

\subsection{Ablation Study}
To better understand the contribution of individual components in our proposed attack, we perform an extensive ablation study. 
This analysis systematically isolates and evaluates the impact of key design choices, including the trace representation, model architecture, and various hyperparameter settings. 
By examining the performance variations with and without specific features or configurations, we aim to provide a comprehensive understanding of their roles in achieving the overall effectiveness and robustness of \attack.

\textbf{Feature and backbone.}
\begin{table}[]
\centering
\caption{Closed-world accuracy (\%) with different settings. Both the IAT histogram and proposed new backbone contribute to the overall performance of \attack.}
\label{tab:abation-study}
\resizebox{\columnwidth}{!}{%
\begin{tabular}{@{}ccccc@{}}
\toprule
\multirow{2}{*}{\textbf{Setting}} & \multicolumn{4}{c}{\textbf{Dataset}}                              \\
                                  & Undefended     & FRONT          & RegulaTor      & Surakav        \\ \midrule
IAT + RF                          & 88.62          & 87.60          & 40.81          & 36.31          \\
TAM + CNN                         & 94.13          & 88.40          & 47.61          & 49.83          \\
\attack                           & \textbf{94.47} & \textbf{93.18} & \textbf{47.78} & \textbf{59.12} \\ \bottomrule
\end{tabular}%
}
\end{table}
We first validate the effectiveness of our proposed feature representation and backbone by considering the following three settings:
\ding{182} RF's backbone with our proposed IAT histogram as input feature (IAT + RF);
\ding{183} our proposed CNN backbone with TAM as input feature (TAM + CNN);
\ding{184} our current design (\attack).
We verify the components on four different datasets: undefended, FRONT, RegulaTor, and Surakav, which represent three different levels of noise magnitude.
The results are presented in Table~\ref{tab:abation-study}.

From the table, we observe that each individual component contributes significantly to performance improvement.
For instance, replacing RF's backbone with our proposed new backbone (\ding{182} $\rightarrow$ \ding{184}) yields notable gains across all datasets, with improvements ranging from 6\% on FRONT to 23\% on Surakav.
Similarly, switching TAM to the IAT histogram (\ding{183} $\rightarrow$ \ding{184}) enhances attack performance, with gains ranging from 0.3\% on the undefended dataset to 9\% on Surakav.
We also observe that the contribution of both components increases as the defense becomes stronger, suggesting that our attack effectively captures key features in challenging scenarios.
Finally, \attack consistently outperforms the individual components, validating the integration of these design choices in achieving state-of-the-art performance.

\textbf{Impact of $G$ and $K$.}
\begin{figure}
    \centering
    \includegraphics[width=\linewidth]{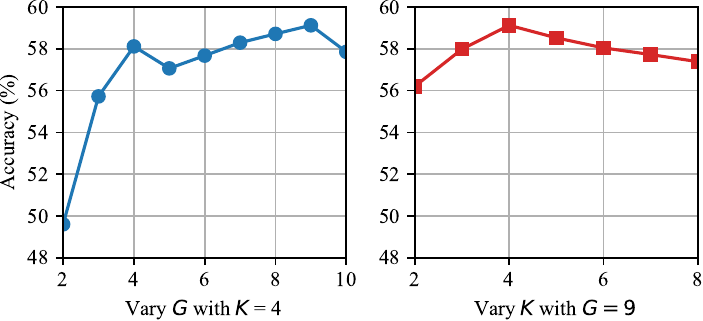}
    \caption{Closed-world accuracy of \attack against Surakav with varying $G$ (left) and varying $K$ (right).}
    \label{fig:vary-g-and-k}
\end{figure}
We next investigate the impact of the bin number $G$ and the kernel number $K$ on \attack’s performance.
The experiments are conducted on the Surakav dataset, as Surakav represents one of the strongest defenses among all evaluated methods.

The parameter $G$ determines the number of bins used in the IAT histogram (see Section~\ref{sec:a-new-trace-representation}).
First, we fix $K=4$ (our default setting) and vary $G$ from 2 to 10.
As shown in Figure~\ref{fig:vary-g-and-k}, the accuracy of \attack increases significantly from 49\% to 58\% as $G$ increases from 2 to 4.
Beyond this point, the accuracy fluctuates slightly around 58\%, peaking at 59\% when $G = 9$.
Based on these results, we set $G = 9$ as the default value.

Next, we fix $G = 9$ and vary $K$ from 2 to 8.
The parameter $K$ specifies the number of kernels used in the Inception2d and Inception1d blocks (refer to Section~\ref{sec:model-architecture}).
While a larger $K$ increases the model’s complexity by adding more kernels, it does not always lead to improved performance.
As shown in Figure~\ref{fig:vary-g-and-k}, when $K$ increases from 2 to 8, the accuracy initially rises from 56\% to 59\%, before dropping to 57\%.
The optimal performance of \attack is achieved at $K = 4$, which we adopt as the default setting.

\section{Discussion}
\label{sec:discussion}

Our proposed attack, \attack, addresses the inherent limitations of existing WF attacks by introducing a novel feature representation, the Inter-Arrival Time (IAT) histogram, and a customized CNN-based architecture.
This section discusses the implications of our findings, the broader impact of this work, and potential defense strategies.

\textbf{Implications of \attack's performance.}
\attack's robustness against modern defenses, including Surakav~\cite{gong2022surakav} and RegulaTor~\cite{HollandH22}, demonstrates the enduring vulnerability of timing-based patterns in network traffic.
Our results suggest that even sophisticated defense mechanisms inadvertently leak information that can be exploited through innovative feature engineering and advanced model design.
Notably, the attack's performance is significantly enhanced by its ability to capture intermediate-grained timing correlations, highlighting the critical role of timing information in WF attacks.

\textbf{Challenges and limitations.}
Despite its robustness, \attack exhibits sensitivity to network conditions, as demonstrated in our experiments with traces from varying circuit bandwidths.
This limitation underscores the importance of comprehensive training datasets that capture diverse network environments.
Moreover, a few works have explored how to strengthen WF attacks under such conditions using data augmentation~\cite{BahramaliBH23, XieCDX00SZ23}.
Future work may consider combine these techniques to enhance the performance of \attack. 

\textbf{Potential mitigation strategies.}
Our work also underscores the necessity of advancing defense mechanisms to counter timing-based attacks effectively.
Potential strategies include: 
\ding{182} \textbf{Combination of Defenses:}
Utilizing TrafficSliver to split traffic and applying additional obfuscation methods (e.g., FRONT~\cite{GongW20} or Surakav~\cite{gong2022surakav}) to each sub-trace could amplify the defense’s resilience.
\ding{183} \textbf{Adversarial Perturbations:}
Introducing noise specifically designed to mislead deep learning models could be an avenue, though practical challenges like real-time implementation must be addressed.
\ding{184} \textbf{Dynamic Traffic Shaping:}
    Designing defenses that dynamically alter traffic patterns in unpredictable ways may thwart timing-sensitive attacks like \attack.

\section{Conclusion}
\label{sec:conclusion}
In this work, we introduced \attack, a novel WF attack leveraging the IAT histogram and an advanced CNN-based architecture.
Through extensive experiments, \attack demonstrated its ability to exploit timing correlations in network traffic, outperforming state-of-the-art attacks against both undefended and defended scenarios.
Notably, our attack achieved 59\% accuracy against Surakav and consistently outperformed other methods across closed- and open-world scenarios.
However, our findings also highlight critical challenges, such as the attack's sensitivity to network conditions and its limited success against deterministic defenses like Tamaraw.
Addressing these challenges requires both enhancing the robustness of WF attacks and developing more sophisticated defense mechanisms.
By advancing the understanding of WF vulnerabilities, this work aims to contribute to the development of more resilient anonymity-preserving technologies.
Future research will explore combining multiple defenses, leveraging adversarial perturbations, and expanding the scope of WF attacks to include evolving network conditions and application scenarios.


\bibliographystyle{plain}
\bibliography{sample-base}

\end{document}